\documentclass[12pt,a4paper]{article}

\usepackage{graphicx}
\usepackage{amsmath,amssymb,mathrsfs}
\usepackage{cite}
\usepackage{bm}

\usepackage{color}
\usepackage{hyperref}
\usepackage{enumerate}
\usepackage{verbatim}

\makeatletter
\newcommand{\rsum}{\DOTSB\rsum@\slimits@}
\newcommand{\rsum@}{\mathop{\mathpalette\rsum@@\relax}}
\newcommand{\rsum@@}[2]{\reflectbox{$\m@th#1\sum@$}}
\makeatother

\usepackage{setspace}
\date{\today}

\def\be{\begin{equation}}
\def\ee{\end{equation}}

\title{\vspace{-1.8cm}Violation of Unitarity in Gravitational Subregions}
\author{
{\Large Aron C. Wall$\,$\footnote{aroncwall@gmail.com; 212 Wellbrook Way, Girton, Cambridgeshire, CB3 0GJ, United Kingdom.}} \\
{\normalsize \vspace{-2pt}Department of Applied Mathematics and Theoretical Physics} \\
{\normalsize University of Cambridge} 
}
\date{\today}

\begin{document}

\maketitle
\vspace{-.8cm}

\begin{abstract}
\noindent This essay contends that in quantum gravity, some spatial regions do not admit a unitary Hilbert space.  Because the gravitational path integral spontaneously breaks CPT symmetry, ``states'' with negative probability can be identified on either side of trapped surfaces.  I argue that these negative norm states are tolerable, by analogy to quantum mechanics.  This viewpoint suggests a resolution of the firewall paradox, similar to black hole complementarity.  Implications for cosmology are briefly discussed.
\end{abstract}

\begin{center}
\fbox{\begin{minipage}{20.8em}
\begin{center}
{\it \small Essay written for the Gravity Research Foundation \newline
2021 Awards for Essays on Gravitation}
\end{center}
\end{minipage}}
\end{center}

\vspace{7pt}

\section{Introduction}
In order to understand a complicated problem, it is usually easiest to break it into simpler pieces.  So it is fortunate that Nature seems to be described by local field theories, where it is possible to study the behavior of fields inside a given spatial region $R$, without reference to the rest of the Universe.

In a quantum field theory, one would like to say that the Hilbert space of the total system factorizes into the Hilbert spaces for $R$ and its complementary region $\overline{R}$:\footnote{Ignoring infrared and ultraviolet divergences.
More rigorously, one uses operator algebras \cite{Haag}.}
\be
{\cal H} \subseteq {\cal H}_R \otimes {\cal H}_{\overline R}.
\ee
Here the subset symbol $\subseteq$ acknowledges that in a gauge theory, ${\cal H}_R$ and ${\cal H}_{\overline R}$ need to be extended to include additional ``edge mode'' degrees of freedom
\cite{Carlip:1996yb,Donnelly:2011hn,Casini:2013rba,Donnelly:2014fua}, due to breaking of gauge-invariance.  

In gravity, after restricting our attention to a region $R$ (ignoring the rest of the Universe), we can no longer consider coordinate changes that move its boundary $\partial R$.  And some diffeomorphisms at $\partial R$ get promoted to ordinary symmetries \cite{Donnelly:2016auv}.

Below, I'll try to convince you that something even worse happens for gravitational subsystems.  Namely, for certain boundaries the decomposition is
\be\label{HV}
{\cal H} \subset {\cal V}_R \otimes {\cal V}_{\overline R}.
\ee
where the ${\cal V}$'s are not Hilbert spaces because they have \emph{negative norm} states $\Psi \in {\cal V}$, with $\langle \Psi\,|\,\Psi\rangle < 0$.

Such states would have negative probability, and therefore one cannot sensibly make predictions for the physics in such regions, taken in isolation.  Predictions can only be made if you consider the entire system, or else restrict attention to a special class of regions $R$ for which this problem does not arise.

This may seem frightening.  But we can make an analogy to the interpretation of quantum mechanics.  It is difficult to assign a sensible physical meaning to the complex amplitudes associated with individual histories in a quantum process, since they interfere.  Only the probabilities of the final outcomes are externally measurable.  In the same way, negative probabilities within subregions might be acceptable, as long as they don't contaminate the experimental predictions of any actual observer.\footnote{In ordinary quantum mechanics, Wigner distributions (over position \emph{and} momentum) also involve negative probability \cite{Wigner:1932eb}.}

\section{The Hartle-Hawking amplitude}

Fig. \ref{f1} shows a Schwarzschild black hole, enclosed between two Anti-de Sitter timelike boundaries:

\begin{figure}[ht]
\centering
\includegraphics[width=.6\textwidth]{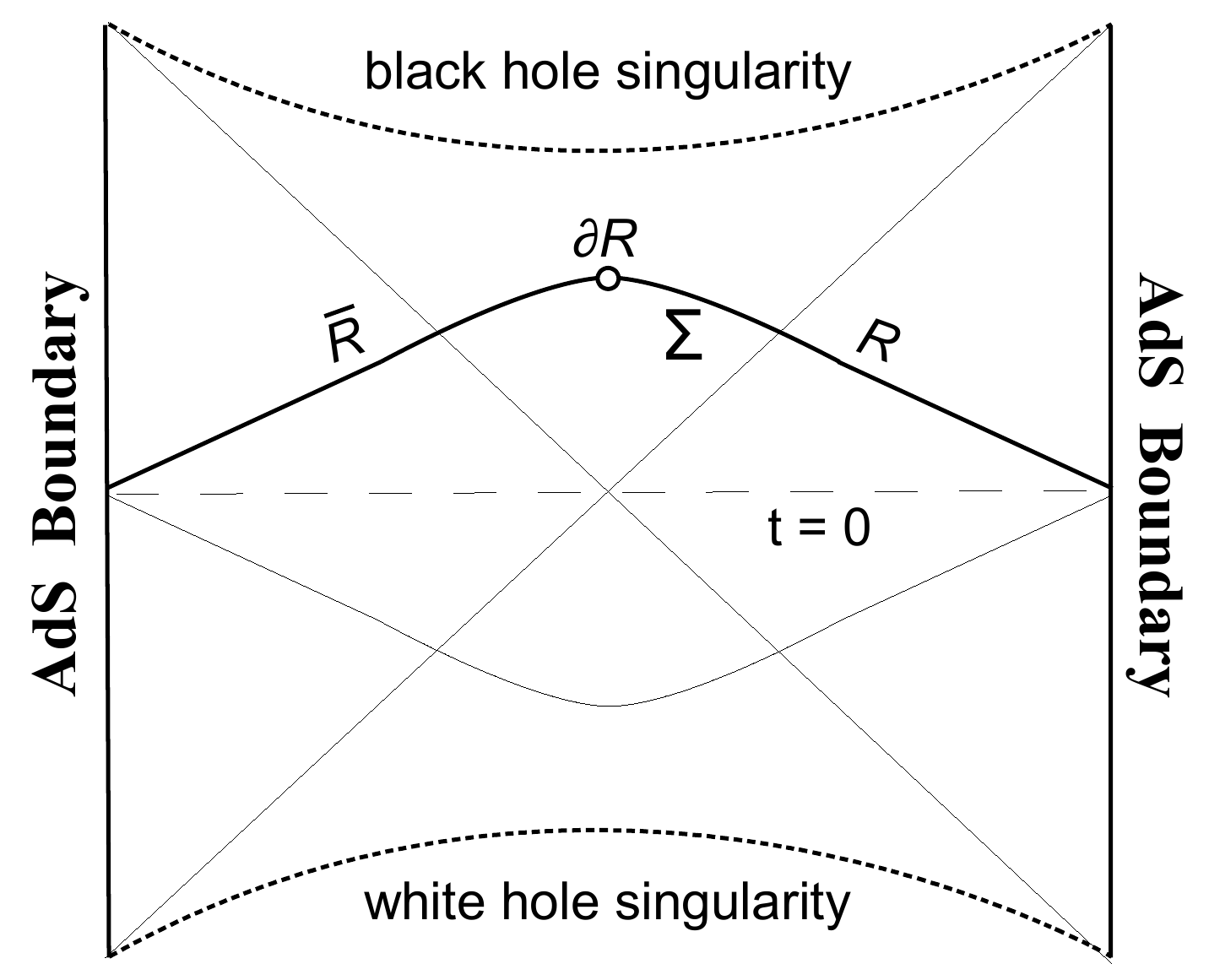}
\caption{\footnotesize The Penrose diagram of AdS-Schwarzschild, with a Cauchy slice $\Sigma$ shown broken into two symmetrical parts $R$ and $\overline{R}$.  The time-reverse of $\Sigma$ (reflected across $t = 0$) is also shown.}\label{f1}
\end{figure}

Pick any smooth Cauchy slice $\Sigma$ which passes through the black hole region, and which intersects each AdS boundary at $t = 0$.  Your choice of $\Sigma$ should be symmetric under left-right reflection.  Cut $\Sigma$ into two pieces, $R$ and $\overline{R}$, at the fixed point of this reflection symmetry (which is therefore $\partial R$).  Let $q_{ij}$ be the spatial metric of $\Sigma$.\pagebreak

So far all this is classical, so let's get more ambitious and ask: what is the amplitude to produce the geometry $q_{ij}$ in the \emph{quantum} theory?  Now the geometry fluctuates, requiring us to choose a specific quantum state, for example the Hartle-Hawking vacuum $\Psi_\text{HH}$.

In the semiclassical limit of large action (${\cal S} \gg \hbar$), the leading-order contribution to $\Psi_\text{HH}$ is determined by the action of a classical solution.  This geometry has an ``imaginary-time'' piece with Euclidean metric signature $(+,+,+\ldots)$, and a Lorentzian piece with metric signature $(-,+,+\ldots)$, joined together at the $t = 0$ slice (Fig. \ref{f2}):
\begin{figure}[ht]
\centering
\includegraphics[width=.5\textwidth]{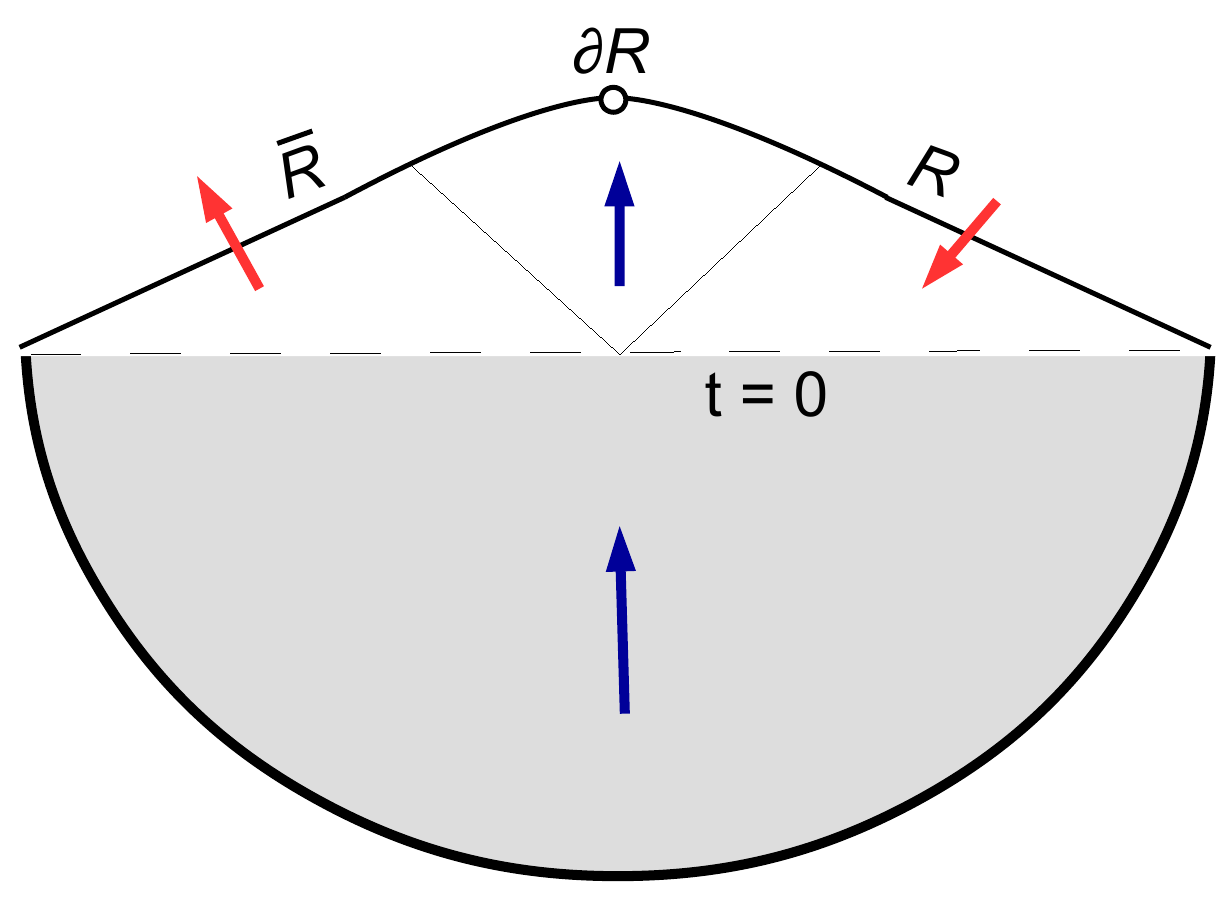}
\caption{\footnotesize The dominant classical solution.  The Euclidean geometry is grey, and the Lorentzian geometry (a portion of Fig. \ref{f1}) is white.  The blue arrow shows the Hartle-Hawking interpretation; the red arrow reinterprets the geometry as a transition from $R$ to $\overline{R}$, via a half-circle imaginary-time evolution of the boundary.} \label{f2}
\end{figure}

In this semiclassical approximation, the amplitude from this geometry is given by:
\be
A = \Delta_\text{loop} e^{i{\cal S}_L/\hbar} e^{-{\cal S}_E/\hbar},
\ee
where ${\cal S}_L$ is the Lorentzian action, ${\cal S}_E$ is the Euclidean action, and $\Delta_\text{loop}$ is a factor coming from small quantum fluctuations, which will not be important.\footnote{Because $\Delta_\text{loop}$ isn't rapidly oscillating.}

Because AdS-Schwarzschild has time-reversal symmetry, there is another equally valid solution, in which the Lorentzian time evolution goes to the past, into the white hole region.  In other words, the Hartle-Hawking wavefunction exhibits \emph{spontaneous breaking of time-reversal symmetry}\footnote{Or in a more general particle physics model, CPT symmetry.} in the Lorentzian regime.  (There is also an unbroken phase, corresponding to Euclidean signature.)

This time-reversed solution contributes the complex-conjugate amplitude $A^*$, ensuring that the final answer is real:
\begin{eqnarray}\label{cos}
\langle q_{ij} | \text{HH} \rangle &\:=\:& A + A^* \:=\: 2\text{Re}(A) \nonumber \\
&\:\sim\:& \cos(-{\cal S}_L/\hbar) e^{-{\cal S}_E/\hbar}.
\end{eqnarray}
But it is not in general positive, due to the rapid oscillation of the cosine if we vary our choice of $\Sigma$ and hence ${\cal S}_L$.  This will imply a violation of unitarity, as we'll see momentarily.

\section{A State with Negative Norm}

Quantum mechanical amplitudes possess a ``crossing symmetry'': meaning that incoming matter particles must be treated similarly to outgoing matter particles.  Analogously, in a gravitational path integral, it shouldn't matter which boundaries we label as ``initial data'' and which as ``final data'' \cite{Reisenberger:1996pu}.

Exploiting this ambiguity, we can reinterpret the same amplitude in another way, as evolution from the region $R$ to the complementary region $\overline{R}$:\footnote{By crossing symmetry, there's no corner term \cite{Hayward:1993my} at $\partial R$, since $\Sigma$ is smooth.  This corresponds to normalizing the inner product so that $\langle q_{ij}^{R} \,|\,q_{ij}^{R} \rangle = \exp({\text{Area}[\partial R]/4})$, which looks suspiciously like a Bekenstein-Hawking state-counting factor \cite{Takayanagi:2019tvn}.}
\be\label{qq}
\langle q_{ij}^{R} \,|\, e^{-\pi H_\text{ADM}/\hbar} \,|\,q_{ij}^{R} \rangle
= A + A^*.
\ee
Let's unpack this expression a little bit: The initial state $|\,q_{ij}^{R} \rangle \in {\cal V}_R$ is a basis state of the spatial metric restricted to $R$.  By reflection symmetry, the metric on $\overline{R}$ is the same as that of $R$, which is why we wrote $\langle q_{ij}^{R} \,|$ for the final state as well.  Finally, the insertion of $e^{-\pi H_\text{ADM}/\hbar}$ corresponds to an imaginary-time evolution on the AdS boundary of size $\pi$ (half a circle), $H_\text{ADM}$ being the energy at the AdS boundary of $R$\cite{Arnowitt:1962hi}.  Since this path integral has the same boundary conditions as shown in Fig. \ref{f2}, its amplitude is the same.\footnote{In particular, if some other geometry were the dominant saddle, then $|\text{HH}\rangle$ would not be close to AdS-Schwarzschild, which would imply that our understanding of black hole thermodynamics is quite badly wrong.
}

But Eq. \eqref{qq} is just the norm of another state, namely $|\Phi^R\rangle \equiv e^{-(\pi/2) H_\text{ADM}/\hbar}|\,q_{ij}^{R}\rangle$, which also lies in the state space ${\cal V}_R$ because energy is gravitationally measurable at the AdS boundary \cite{Marolf:2008mf}.  Yet because of the rapidly oscillating cosine in Eq. \eqref{cos}, this norm isn't always positive:
\be
\langle \Phi^R \,|\, \Phi^R \rangle \ngeq 0,
\ee
where we get the negative sign about 50\% of the time, if we select $\Sigma$ carelessly.  Thus we find that the quantum gravity theory restricted to one side of a surface $\partial R$ cannot always be unitary!


\section{Unitarity and Firewalls}

To interpret this result, it will be helpful to review three entirely distinct (mix-and-match) senses of unitarity.  Non-unitarity could mean:
\begin{description}
\item (i) The state space ${\cal V}$ does not have a positive semi-definite norm with $\langle \Psi\,|\,\Psi\rangle \ge 0$ for all vectors $\Psi \in {\cal V}$.\footnote{Vectors with zero norm are no big deal, since you can still construct a Hilbert space by quotienting them out, using gauge-invariance.}

\item (ii) The norm on ${\cal V}$ is not preserved by (Lorentzian) time evolution, i.e. the Hamiltonian is not self-adjoint.

\item (iii) Pure states evolve to mixed states (represented by density matrices 
$\rho \in {\cal V} \otimes {\cal V}^*$).
\end{description}
The famous ``black hole information paradox'' concerns whether black hole formation and evaporation is unitary in sense (iii), whereas we have been discussing sense (i).

There is, however, a potential application to the information puzzle.  Black hole unitarity leads to danger from the ``firewall paradox'' \cite{Mathur:2009hf,Almheiri:2012rt,Braunstein:2009my}, in which a near-vacuum horizon is incompatible with quantum mechanics.  

The problem arises from the Strong Subadditivity inequality: for any 3 disjoint quantum systems, the von Neumann entropy $S = \text{tr}(\rho \ln \rho^{-1})$ satisfies
\be\label{SSA}
S(AB) + S(BC) \ge S(A) + S(C).
\ee
This inequality leads to a contradiction for an evaporating black hole, as shown in Fig. \ref{f3}:
\begin{figure}[ht]
\centering
\includegraphics[width=.39\textwidth]{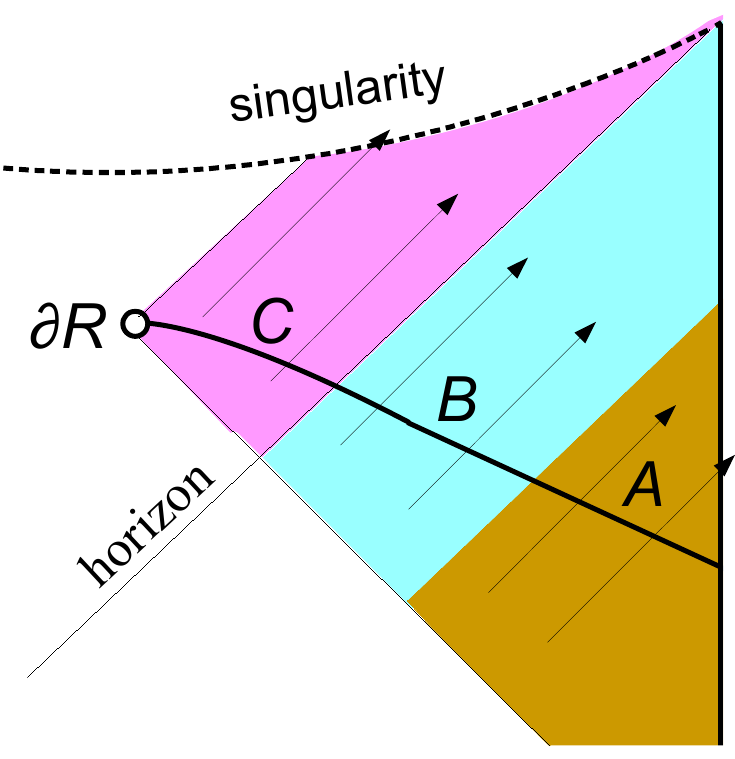}
\caption{\footnotesize A black hole formed from collapse, shown at late times.  $A = $ ``the entire first half of the Hawking radiation'', $B = $ ``some later Hawking modes just outside the horizon'', and $C =$ ``the partner Hawking modes just inside the horizon''.  Unitary evaporation in sense (iii) requires that $S(AB) < S(A)$, while regularity at the horizon requires that $C$ to purify $B$ so that $S(BC) < S(C)$, violating Strong Subadditivity \eqref{SSA}.  But if unitarity (i) is violated in $R = ABC$, all bets are off.} \label{f3}
\end{figure}

But quantum inequalities are only valid under the assumption of a positive norm, and hence positive density matrices.  Hence, violations of unitarity (i) might save us from violations of unitary in sense (iii).\footnote{For fans of the AdS/CFT holographic correspondence: violations of unitarity (i) on the AdS side wouldn't imply violations in the dual field theory, because only ``entanglement wedges'' are dual to CFT regions \cite{Jafferis:2015del,Dong:2016eik,Cotler:2017erl}.}  

The idea is similar to the ``black hole complementarity'' picture \cite{Susskind:1993if}, but \emph{without} the assumption that the region causally accessible to a complete worldline has a unitary Hilbert space.  (Thus potentially evading the firewall paradox.)

\section{Implications for Cosmology}

Our own Universe also breaks time-reversal symmetry, because it is expanding from an initial singularity.  Does this also signal a breakdown of unitarity?  If so, then we can't calculate probabilities in such regions, except as an intermediate step in a longer spacetime history.

In the black hole case, the violation occurs when $\partial R$ is a ``trapped surface'', from which lightrays shot in both normal directions have contracting area.  So it is reasonable to expect that violations of unitarity in cosmology will also occur for the time-reverse case: anti-trapped surfaces (whose area is expanding in both null directions).

This could justify the hypothesis that the most natural place to define a Hilbert space in cosmology is a ``holographic screen'' \cite{Bousso:1999cb,Nomura:2016ikr}, which is just barely not anti-trapped.  This screen would govern all the interior dynamics, almost like a modern reincarnation of Ptolemy's ``sphere of fixed stars'' \cite{Ptolemy}.  

Anti-trapped surfaces \emph{do} exist within our past lightcone, so we can't just ignore these nonunitary regions.   Nevertheless, as long as we deploy the QM Born rule \emph{only} in a region $R$ with a true Hilbert space ${\cal H}_R$, the probabilities will be positive.  And as long as unitarity (ii)---that the norm is preserved by time evolution---is still sacred, these probabilities will continue to add to 1 (assuming the initial state is properly normalized).

Yet perhaps the precise distribution of statistical measurements in $R$ would still be inexplicable on the assumption of a unitary quantum history; just as the statistics of quantum entanglement cannot be duplicated by any classical process \cite{Bell:1964kc}.  It would be interesting to check implications for models of inflation.  

In any case, unless these negative probabilities can be banished, quantum gravity forces us into a new epistemology, even subtler than what was needed for the atom.

\subsubsection*{Acknowledgements}\vspace{-4pt}
{\footnotesize I am grateful for support by AFOSR grant ``Tensor Networks and Holographic Spacetime'', STFC grant ``Particles, Fields and Extended Objects'', and conversations with Ted Jacobson, Raphael Bousso, William Donnelly, Amr Ahmadain, Rifath Khan, Gon\c{c}alo Araujo-Regado, Don Marolf, Ahmed Almheiri, Laurent Friedel, Gary Horowitz, Steve Carlip, Adam Brown, and Daniel Jafferis.}

\end{document}